Igor O. Zavadskyi (Taras Shevchenko National University of Kyiv)
ihorza@gmail.com

# A family of fast exact pattern matching algorithms

**Abstract.** A family of comparison-based exact pattern matching algorithms is described. They utilize multi-dimensional arrays in order to process more than one adjacent text window in each iteration of the search cycle. This approach leads to a lower average time complexity by the cost of space. The algorithms of this family perform well for short patterns and middle size alphabets. In such case the shift of the window by several pattern lengths at once is quite probable, which is the main factor of algorithm success. Our algorithms outperform the Boyer-Moore-Horspool algorithm, either in the original version or with Sunday's "Quick search" modification, in a wide area of pattern length / alphabet size plane. In some subareas the proposed algorithms are the fastest among all known exact pattern matching algorithms. Namely, they perform best when alphabet size is about 30–40 and pattern length is about 4–10. Such parameters are typical for search in natural language text databases.

## 1 Introduction

Pattern matching is one of the most fundamental techniques used in computer science. The most common pattern matching problem is formulated as finding all the exact occurrences of a given substring in a larger body of text. Through the entire presentation we use the following notation:

$T$ – input text
$P$ – pattern to be searched
$n$ – length of the input text
$m$ – length of the pattern
$\Sigma$ – alphabet of input text and pattern
$|\Sigma|$ – size of alphabet $\Sigma$
$|\Sigma_P|$ – number of different symbols in the pattern

This problem has been systematically studied since the beginning of seventies. The number of algorithms more efficient than the simplest straight forward search (SF) has been discovered. The most famous of them are the Knuth-Morris-Pratt algorithm [1], which improves the worst-case time complexity of the SF from $O(nm)$ to $O(n+m)$ and Boyer-Moore algorithm [2], which significantly outperforms the SF in the average.

Almost all known pattern matching algorithms include the preprocessing stage, when some preliminary values are obtained basing on the pattern, and the main search cycle, when the text body is scanned. In most cases the algorithm efficiency strongly depends not only on the pattern and text lengths, but also on the alphabet size. Since, as a rule, the dependence of the main search cycle on the text length is linear and time complexity of the preprocessing stage is negligibly small, it is worthwhile to compare the algorithm efficiency on the $(|\Sigma|,m)$-plane.

Our research concerns the left up area, where $m$ is small, $|\Sigma|$ is large. In this area two modifications of BM algorithm, namely Boyer-Moore-Horspool algorithm (BMH) [3] and Sunday's "Quick Search" (QS) [4] were considered the best for a decades. However, in 2000s a number of more efficient exact pattern matching algorithms were invented. According to [5] they are FJS for $m<8$ and $|\Sigma|>32$, TVSBS for $m=2$ and $8\leq|\Sigma|\leq32$, EBOM for $4\leq m\leq16$ and $8\leq|\Sigma|\leq32$, SBNDM for $8\leq m\leq16$ and $|\Sigma|>32$ and FSBNDM for $m>16$ and $|\Sigma|>8$. The algorithms mentioned cover all three known types of pattern-matching algorithms: FJS and TVSBS are comparison-based; EBOM is automata based, while SBNDM and FSBNDM algorithms utilize the bit-parallel operations. We propose the new comparison-based algorithms. Almost all algorithms of this type, including FJS, TVSBS and our new algorithms, exploit the idea of bad-character shift, which originates from BM search. It is to compare the last character of the search window and the last character of the pattern and, if they do not match, shift the window as long as possible. The BMH algorithm is based on this idea only. We develop a generalization of BMH algorithm, which allows performing several bad-character shifts at each iteration.

The search cycle of the BMH algorithm is shown in Fig. 1. Since we analyze the algorithms in operation level, we try to remove the unnecessary subtractions denoting $n–m$ by $nm$ and $m–1$ by $m1$; these values can be calculated in a preprocessing stage. The bad character shift is performed in the row 7 and its length is equal to $D[T[pos+m–1]]$, where $pos$ is the current position of the search window and $D$ is the shift array calculated in a preprocessing stage. If the ratio $|\Sigma_P|/|\Sigma|$ is small enough, the symbol $T[pos+m–1]$ most likely does not occur in the pattern and the length of this shift is maximum, i.e. it stands *m*. These maximum length shifts are the main factor responsible for the efficiency of BMH in the left up area of the ($|\Sigma|$,*m*)-plane. And we are interested in the most left upper subarea, where the ratio $|\Sigma_P|/|\Sigma|$ is particularly small. In this case one can assume that probably not only character *T*[*pos*+*m*–1] does not belong to the pattern, but the characters *T*[*pos*+2*m*–1], *T*[*pos*+3*m*–1] etc. as well. This means that the search window can be shifted by several window lengths at once, or, in other words, several adjacent search windows can be processed at iteration of the search cycle. This is the main idea of the multi-window search algorithms.

```
1.  while pos<nm
2.      j ← m1;
3.      while j>0 AND T[pos+j]=P[j]
4.          j ← j-1;
5.      if j=0
6.          output pos;
7.      pos ← pos+D[T[pos+m1]];
```

Fig. 1. The main search cycle of the Boyer-Moore-Horspool algorithm

Of course, at least *k* symbols of input text must be read and processed in each substring of the length *km* in order not to miss the possible pattern occurrence. Thus, at least *k* readings of input text characters should be done for each window of the length *km* – just the same number as at *k* iterations of the single-window algorithm like BMH or QS. However, we can reduce the number of other operations using the *k*-dimensional arrays. Such arrays occupy rather more memory than the pattern shift arrays for single-window search algorithms and their filling takes more preprocessing time. Nevertheless, as will be shown below, these space overheads are not that big comparing to memory size of modern computers, while time overheads are more than covered in the main search cycle, resulting in a significant gain in the total for the wide range of alphabet size / pattern length combinations.

The idea of using two text windows while searching all occurrences of the pattern is not new. It was utilized in Two-Sliding-Windows (TSW) algorithm [6]. However, these windows were supposed to be processed in parallel and thus TSW algorithm is suitable for parallel processor structures only. Also the idea of two-dimensional search array was already implemented in a number of algorithms, for instance, in Berry-Ravindran algorithm [7], TVSBS and EBOM. However, it was always proposed to use two adjacent characters of a text as indices. This significantly increases the probability of the maximum length shift if it is low for single-character check but otherwise leads to superfluous density of the checks. In other words, if even single-character check causes the maximum shift with high probability, there is no need to check two adjacent characters to shift the text window by *m* or *m*+1 positions. In this case it may be better to perform double-check of the characters `T[pos]` and `T[pos+m-1]`, which could shift the text window by 2*m* positions at once.

## 2  Double window algorithm

Let us discuss how to process two adjacent search windows of the length *m*, which could be considered as one window of the double length 2*m*. We try to reduce the total number of computing operations required to process the substring of the length 2*m*. Let us examine the main cycle of BMH algorithm shown in Fig. 1. Two reads from shift table *D* in row 7 in two iterations can be replaced by one if we use two-dimensional shift table $D2_{|\Sigma|\times|\Sigma|}$ defined as follows.

*D*2[*i*][*j*] is the leftmost possible position of the first character of the pattern under assumption that *T*[*m*–1]=*i* and *T*[2*m*–1]=*j*. All shifts defined by the table *D*2 can be divided in 4 types shown in Fig. 2.

a)  Neither *i* nor *j* belong to pattern *P*. Then *P* can be safely shifted by 2*m* positions forward.

b)  Character *i* doesn't belong to pattern *P* but *j* belongs. In this case *P* can be safely shifted by more than *m*–1 symbols but less than 2*m*. Namely, the rightmost occurrence of *j* in *P* should be aligned with T[2*m*–1].

c)  Character *i* belongs to *P* and *P*[*m*–1]≠*i*. Then *P* can be safely shifted forward by less than *m* symbols. Namely, the rightmost occurrence of *i* in *P* should be aligned with *T*[*m*–1].

d)  *P*[*m*–1]=*i*. This is the case when the pattern can be matched in the current position. One should check if *T*[0]…*T*[*m*–1] coincides with the pattern before proceed forward.

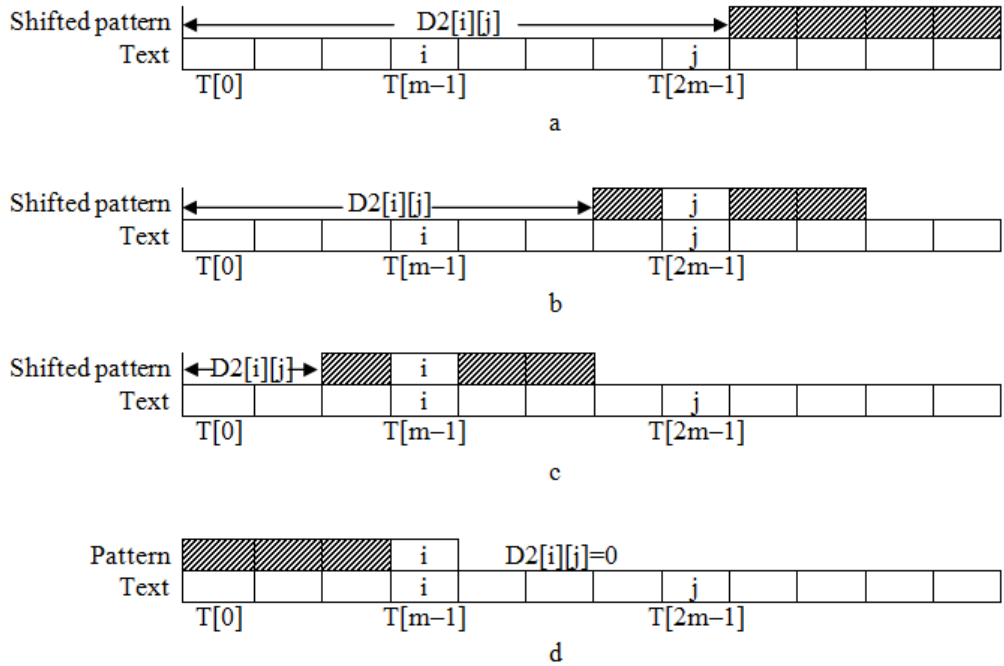

Fig. 2. Pattern shifts in Double window algorithm

```
1.      while pos<nm
2.          r ← D2[T[pos]][T[pos+m]];
3.          if r=0
4.              j ← 0;
5.              while j<m AND T[pos-m1+j]=P[j]
6.                  j ← j+1;
7.              if k=m
8.                  output pos-m1;
9.              pos ← pos+D[T[pos]];
          else
10.             pos ← pos+r;
```

Fig. 3. The Double window algorithm main search cycle

Let us calculate the number of operations in BMH and Double window (DW) algorithms required for the shift of the text window by $2m$ characters forward in the case of the maximum possible shift. This is the most probable case when pattern length is very small compared to alphabet size. In this case only rows 1, 2, 3 and 7 in two iterations of BMH algorithm and only rows 1, 2, 3 and 10 in one iteration of Double window algorithm are executed. Note that getting the element of one-dimensional array like $D[x]$ is equivalent to $*(D+x)$ in C notation, which requires one addition and two readings from memory, while getting the element of two-dimensional $|\Sigma|\times|\Sigma|$ array like $D2[x][y]$ is equivalent to $*(D2+|\Sigma|*x+y)$, which requires two additions, one multiplication and three readings from memory ($|\Sigma|$ is a constant).

The calculations are shown in Table 1. The number of operations in Double window algorithm is more than twice less compared to BMH. Also, it should be noted that memory reads generally take longer time than operations that could be performed using processor registers only, e.g. arithmetic operations. If someone implements the discussed algorithms in assembler language, the values of variables $pos$, $m1$, $m$, $j$ etc. can be stored in processor registers, while the arrays $T$, $D$ and $D2$ cannot. In such implementation 8 memory reads are performed in two iterations of BMH (2 in row 3 and 2 in row 7 in each iteration) and only 3 memory reads needed in one iteration of Double window algorithm, they are $T[pos]$, $T[pos+m]$ and $D2[T[pos]][T[pos+m]]$ in row 3. Thus, in the case of efficient assembler implementation, the Double window algorithm main cycle outperforms the

BMH main cycle even more than in the case of implementation in the high-level programming language, if the longest possible shifts are performed (Figure 2a).

Table 1. The operational complexity of Boyer-Moore-Horspool and Double window algorithms

| Operation | Boyer-Moore-Horspool | Double window |
|---|---|---|
| comparison | 4 – (rows 1 and 3)x2 | 2 (rows 1 and 3) |
| assignment | 4 – (rows 2 and 7)x2 | 2 (rows 2 and 10) |
| memory reads | 28 – (2 in row 1; 1 in row 2; 6 in row 3; 5 in row 7)x2 | 10 (2 in row 1; 6 in row 2; 2 in row 10) |
| additions | 14 – (3 in row 3; 4 in row 7)x2 | 6 (5 in row 2 and 1 in row 10) |
| multiplication | – | 1 (row 2) |
| Total | 50 | 21 |

One can observe that even one iteration of Double window algorithm main cycle requires fewer operations than one iteration of BMH main cycle in the case when the condition $r=0$ is not met. Therefore, in the cases shown in figures 2b and 2c the Double window algorithm main cycle still executes faster than BMH main cycle. Note that in the case (c) the equality $D2[i][j]=D[i]$ holds, i.e. the shift length in Double window algorithm is just the same as in BMH.

Of course, the advantage of DW main cycle over BMH main cycle in the case (b) is lower than in the case (a) and in the case (c) is lower than in the case (b). While the ratio $|\Sigma_P|/|\Sigma|$ increases, the balance between cases (a), (b) and (c) moves to (b) and (c) and then to (c) only. If $|\Sigma_P|/|\Sigma|$ is close to 1, the case (c) occurs almost always and outperformance of DW main cycle over BMH main cycle is small.

The case shown in Figure 2d for random text and pattern occurs with the probability $1/|\Sigma|$ regardless of $|\Sigma_P|$ value. In this case the internal cycle of DW in rows 5 and 6 executes and the internal cycle of BMH in rows 3 and 4 has more than one iteration. Each iteration of DW internal cycle requires more time than that one of BMH, because the comparison $T[pos–m1+j]=P[j]$ requires 4 additions/subtractions and 6 readings from memory, while the comparison $T[pos+j]=P[j]$ consists of only 3 additions and 5 readings from memory.

As a result, the main cycle of the DW algorithm is essentially faster than the main cycle of BMH algorithm when the following conditions are met:
 (1) the alphabet is large enough to make the case (d) not frequent;
 (2) the ratio $|\Sigma_P|/|\Sigma|$ is small enough to make the case (c) not frequent.

The simulation shows that only when alphabet size is up to 4, the condition (1) violation forces the DW algorithm main cycle to run slower than BMH main cycle on random pattern and text. Any alphabet of size greater than 4 could be considered as "large enough". The violation of condition (2) forces the DW algorithm main cycle to run approximately at the same speed as the BMH main cycle. However, for the wide range of pattern length / alphabet size combinations the DW is essentially faster than BMH. This range covers $m$ in the range 20–30 and $|\Sigma|>4$.

### 3 Multi-window extension

Let us consider the possibility of processing more than 2 adjacent text windows in one iteration. The modification of the DW algorithm is simple: the $N$-dimensional array $DN$ should be used instead of $D2$. It is defined as follows.

$DN[i_1]…[i_N]$ is the leftmost possible position of the beginning of the pattern under assumption that $T[km–1]=i_k$, $k=1,..,N$.

In Figure 3 only row 2 should be changed in a following way:

  $r \leftarrow DN[T[pos]][T[pos+2m]]…[T[pos+Nm]]$;

Thus we obtain the Triple Window (TW), Quadruple Window (QW) and other Multi-window algorithms.

Using C notation this assignment can be rewritten as $r=*(DN+b_{N-1}pos+…+b_1(pos+(N–1)m)+pos+Nm)$, where $b_k=|\Sigma|^k$. Of course, in order to reduce the number of multiplications, the values $2m,…,Nm$ can be calculated in the preprocessing stage, while the values $b_k$ could be considered as constants. However, every next dimension adds two additions, one multiplication and two memory reads to the calculation routine. This overhead is covered by longer shifts until the value of $|\Sigma_P|/|\Sigma|$ is small enough, but in the case $N≥3$ the preprocessing begins to play an important role, since its time and space complexity grows exponentially depending on $N$. The preprocessing stage is discussed in the next section.

### 4 Preprocessing

On the preprocessing stage of the $N$-window algorithm the values $km$, $k=1,…,N$ are calculated and arrays *D* containing |Σ| elements and *DN* with |Σ|$^n$ elements are filled with the values. Filling the array *D* is just the same as in the BMH algorithm and runs in *O*(|Σ|). Obtaining values *km* is *O*(*N*) time, which is ignorable small for realistic values of *N*. Filling the array *DN* takes up almost all the time. The following procedure completes this task.

1. Assign the value *Nm* to all elements.
2. Replace the values *D*[$i_1$]…[$i_N$], where $i_N \in P$, with *Nm*–$q_N$–1, where $q_N$ is the rightmost position of the character $i_N$ in the pattern *P*.

…

*N*+1. Replace the values *D*[$i_1$]…[$i_N$], where $i_1 \in P$, with *m*–$q_1$–1, where $q_1$ is the rightmost position of the character $i_1$ in the pattern *P*.

The first step takes $O(|\Sigma|^N)$ time, while each other step – $O(m|\Sigma|^{N-1})$ time. The overall time complexity of preprocessing stage is $O(|\Sigma|^N + Nm|\Sigma|^{N-1})$.

Using the special functions that copy memory blocks, like `memcpy` from `memory.h` C library, one can build the implementation that is faster in times than the conventional method given above.

The space complexity of Multi-window algorithms is, of course, strongly greater than that one for BMH/QS. However, memory requirements of *D*2 array even for relatively large alphabet containing 256 symbols are only 64 Kb, which is absolutely admissible for present-day computers and programs.

The TW algorithm, as computational experiment shows, is efficient only for b values around 32. Then the size of D3 (32x32x32) array is only 32 Kb. The QW algorithm could be efficient for b=16 at maximum and the size of the respective D4 array is 64 Kb.

## 5 Unrolling cycle

The algorithms based on bad character rule, such as BMH or Multi-window, could be accelerated using the unrolling cycle technique. It consist of applying "blind" shifts, without checking the end of file, until shift value is not positive. In order not to miss the end of the file, the text is appended by the fictitious pattern.

This technique could be applied to Multi-window algorithms in a following way:
- Row 1 should be replaced with the endless cycle "`while 1`".
- After row 3 we should check the condition *pos*≥*nm* and break the cycle if it is met.

This allows us to check the end of the file only when the shift array element is equal to zero and speed up the multi-window algorithms by 2–4%.

## 6 Computational experiment

We implement the unrolled versions of Double window, Triple-window (TW), Quadruple-window (QW) and a number of other known algorithms in C language, use the Microsoft Visual studio compiler to build the executables and run them on the Athlon II X2 245 processor, 2.9 GHz, Windows 7 platform. Preprocessing stage of the TW and QW algorithms is implemented using the fast memory fill functions. The text containing 10 MB characters is randomly built and the patterns as well. The distribution of characters is uniform.

The results for alphabet of size 32 and different pattern lengths are presented in Table 2. The total running time of 1000 runs is shown. The QW results are not shown, since they are significantly worse than DW and TW results for alphabet size 32. The QW algorithm could be efficient for smaller alphabets and very short patterns.

Table 2. The total running time of pattern matching algorithms, 10 MB of text × 1000, in seconds

|      | BMH    | QS     | DW     | TW         | EBOM   | FJS    | TVSBS      | SBNDM      | FSBNDM     |
|------|--------|--------|--------|------------|--------|--------|------------|------------|------------|
| m=2  | 45,326 | 31,678 | 34,030 | 27,229     | 90,742 | 26,169 | **24,186** | 25,689     | 36,091     |
| m=3  | 33,320 | 25,596 | 24,429 | 21,396     | 52,538 | 22,890 | 21,109     | **20,034** | 26,177     |
| m=4  | 24,552 | 20,366 | 18,836 | **15,812** | 33,695 | 19,420 | 18,182     | 16,071     | 19,477     |
| m=5  | 19,919 | 17,358 | 15,558 | **13,630** | 25,020 | 16,397 | 16,025     | 14,127     | 16,273     |
| m=6  | 16,458 | 14,583 | 13,313 | **11,623** | 19,524 | 13,926 | 13,824     | 12,328     | 13,250     |
| m=7  | 15,049 | 13,724 | 11,957 | **10,706** | 17,003 | 12,880 | 13,002     | 11,855     | 11,930     |
| m=8  | 14,279 | 13,213 | 11,704 | **10,913** | 15,945 | 12,550 | 12,680     | 11,725     | 11,451     |
| m=9  | 12,170 | 11,327 | 10,181 | **9,316**  | 13,299 | 10,846 | 11,212     | 10,391     | 9,624      |
| m=10 | 11,097 | 10,411 | 9,363  | 8,990      | 11,667 | 9,826  | 10,218     | 10,076     | **8,963**  |
| m=11 | 10,541 | 9,788  | 9,088  | 8,547      | 10,673 | 9,175  | 9,454      | 9,401      | **8,103**  |
| m=12 | 9,874  | 9,432  | 8,522  | 8,198      | 9,949  | 8,859  | 9,071      | 9,271      | **7,726**  |

**7   Conclusions**

As is seen, the Triple Window algorithm is superior over all known algorithms for pattern lengths from 4 to 9 and alphabet size 32. Both DW and TW algorithms outperform classical comparison-based algorithms, such as BMH and QS for all considered values of *m*. The TVSBS, FJS and SBNDM are better for very short patterns, while for pattern length 10 and more the FSBNDM algorithm becomes superior.

The technique used in Multi-window algorithm family could be applied in order to accelerate the algorithms based on comparisons of adjacent characters, which is a future research direction.